\documentclass{PoS}
\def\Dslash{\mathop{\not\!\! D}}

\title{Pseudoscalar Flavor-Singlet Physics with Staggered Fermions}

\ShortTitle{Pseudoscalar Flavor-Singlets with Staggered Fermions}

\author{UKQCD Collaboration}

\author{\speaker{Eric B. Gregory}\thanks{University of Glasgow from October 2007}, Alan Irving, Chris M. Richards\\
        Department of Mathematical Sciences, University of Liverpool, 
Liverpool, L69-7ZL, UK\\
        E-mail: \email{e.gregory@physics.gla.ac.uk}}

\author{Craig McNeile\\
Department of Physics and Astronomy,
University of Glasgow, Glasgow G12-8QQ, UK\\
}

\author{Alistair Hart\\
School of Physics (JCMB), University of Edinburgh, Edinburgh. EH9-3JZ, UK\\
}

\abstract{Accurately calculating the mass of flavor-singlet meson states from 
numerical lattice simulations is an important milestone for lattice QCD. 
Careful measurement of the full pseudoscalar flavor-singlet propagator is 
also a crucial step in understanding the dynamics of the fermion sea on the 
lattice, in particular for potentially non-trivial formulations such as with 
2+1-flavor staggered fermions. 
We briefly describe details of a dynamical QCD calculation using
improved staggered fermions, with 30,000 trajectories, that was run
for our studies of flavour singlet mesons.
}

\FullConference{The XXV International Symposium on Lattice Field Theory\\
		 July 30-4 August 2007\\
		 Regensburg, Germany}

\begin{document}

\section{Introduction} 
An first-principles calculation of the mass of the $\eta^\prime$ meson
has long been a goal of lattice quantum-chromodynamics. Such a calculation 
would shed light on the dynamics of the fermionic sea and the mechanism by
which fluctuations raise the mass of the $\eta^\prime$ mass over that of the 
pion \cite{Witten:1979vv, Veneziano:1979ec}.

A number of collaborations have performed simulations of pseudoscalar singlet systems with
$N_f=2$ flavors of dynamical fermions, e.g.\cite{Itoh:1987iy,McNeile:2000hf,
Venkataraman:1997xi,Kogut:1998rh, Fukaya:2004kp}.
Recently the JLQCD/CP-PACS collaboration reported on
a preliminary calculation of the $\eta$ and $\eta'$
mesons with 
$2+1$ flavours of Wilson fermions~\cite{Aoki:2006xk}. This article 
is based on the work with $2+1$ flavors described more fully in 
\cite{Gregory:2007ev}.

There are compelling reasons to attempt such a calculation with improved 
$2+1$ flavors of staggered fermions. Improved staggered fermions
have an impressive track record of accurate hadron spectroscopy of 
flavour non-singlet quantities
with light dynamical quarks~\cite{Davies:2003ik,Aubin:2004wf}, 
suggesting it would be a good testing ground for 
the difficult task of singlet spectroscopy.
The flavor-singlet propagator contains disconnected diagrams, which are 
inherently noisy in numerical simulations and require long Monte Carlo
timeseries to measure with precision. Hence it is imperative that any program
intent on singlet spectroscopy be capable of fast simulation of
dynamical quarks.

The staggered fermion program, however, is nagged by questions about the 
validity of the fourth-root trick which is employed to convert the
native four tastes of staggered fermions to the desired $N_f$ flavors. Recent
theoretical work suggests that in the continuum 
limit~\cite{Durr:2005ax,Sharpe:2006re} problems with the fourth root may 
vanish, but if not, the pseudoscalar flavor-singlet system is a likely
arena to find evidence of any 
inconsistencies~\cite{Creutz:2007yg,Creutz:2007rk}.

With this in mind we have endeavoured to simulate the $\eta/\eta^\prime$
system with $N_f=2+1$ flavors of improved staggered fermions. In this 
article we report on the conclusions of the first stage of this project ---
an attempt at pseudoscalar flavor-singlet spectroscopy on the library
of MILC ``coarse'' lattices, and discuss the prospects for the project in its 
second stage, namely measurement of the pseudoscalar flavor-singlet on a 
timeseries consisting of roughly 30000 trajectories of $2+1$-flavor improved
staggered fermion configurations.

\section{Theoretical background}   \label{se:theory}

In general, the propagator for the pseudoscalar singlet meson with
$N_f$ flavors of quarks is given by
\begin{equation}
G_{SP}(x',x)=\langle\sum_{i=1}^{N_f}\overline{q}_i(x')\gamma_5q_i(x')\sum_{j=1}^{N_f}\overline{q}_j(x)\gamma_5q_j(x)\rangle.
\end{equation}
From the two types of contractions in this expression we get $N_f$ connected
diagrams:
\begin{equation}
\langle\sum_{i}\overbrace{\overline{q}_i(x')\gamma_5\underbrace{q_i(x')\sum_{j}\overline{q}_j}(x)\gamma_5q_j}(x)\rangle,
\end{equation}
and $N_f^2$ disconnected terms:
\begin{equation}
\langle\sum_{i}\overbrace{\overline{q}_i(x')\gamma_5q_i}(x')\sum_{j}\overbrace{\overline{q}_j(x)\gamma_5q_j}(x)\rangle.
\end{equation}
In the $N_f=2$ flavor-symmetric case, the connected term is identical with 
the pion propagator.

So we can write:
\begin{equation}
\label{degenerate_singlet_prop}
G_{SP}(x',x)= N_fC(x',x)-N_f^2D(x',x),
\end{equation}
where the extra fermion loop in the disconnected diagram gives rise to 
the relative minus sign, and
\begin{equation}
G_{NP}(x',x)= N_fC(x',x).
\end{equation}

A significant difference between the connected and disconnected terms is the
role of the sea quarks bubbles in the disconnected term. One way to highlight 
the effects of the sea quarks is to take the ratio of the disconnected to 
connected terms. We expect that at large Euclidean time separations in full 
QCD 
\begin{equation}
G_{NP}(t) = A_{NP}e^{-M_{NP} t}\qquad{\rm and}\qquad G_{SP}(t) = A_{SP}e^{-M_{SP} t}.
\end{equation}
So we would expect the ratio of disconnected to connected parts to go as
\begin{equation}
R(t)=\frac{N_f^2D(t)}{N_fC(t)}=1-\frac{A_{SP}}{A_{NP}}e^{-(M_{SP}-M_{NP}) t}
\label{fullQCD_R}
\end{equation}
in the $SU(N_f)$ flavor-symmetric case. In the relevant $N_f=2+1$ case we 
would instead consider a modified ratio:
\begin{equation}
R(t)_{SU3}  =\frac{4D_{qq}(t) + 4D_{qs}(t)+ D_{ss}(t)}{2C_{qq}(t)+C_{ss}(t)},
\label{su3R}
\end{equation}

By constructing this ratio we can highlight the behaviour of the sea quarks. 
For example, if instead of simulating full QCD we perform a 
quenched simulation, we should 
expect to find
\begin{equation}
R(t)=\frac{D(x',x)}{C(x',x)}= 
\frac{(m_0^2 - \alpha M_{NP}^2) } {2 M_{NP}}t 
+ \frac{m_0^2 + \alpha M_{NP}^2}{ 2 M_{NP}^2},
\label{quenched_R}
\end{equation}
where $\alpha$ is the parameter of the kinetic
term of singlet pseudoscalar meson~\cite{Bernard:1992mk},
and $m_0$ is the difference between the masses $M_{SP}$
and  $M_{NP}$. 

By measuring the $D/C$ ratio in (\ref{su3R}) we might expect to find any 
problems 
with the fermionic sea introduced by the fourth-root trick.

\section{Simulation details}
In the first phase of this project, we measured the pseudoscalar singlet 
connected correlator and the fermionic loops required to construct
the pseudoscalar singlet disconnected correlator on $N_f=0,3$ and $2+1$ 
flavor lattices primarily from MILC's ``coarse'' ensemble library.
The specific ensembles
examined are listed in Table \ref{ensembles}. 

To more fully understand the fluctuations inherent in disconnected correlator 
measurements, we extended the quenched ensemble from 408 configurations to 
6154 configurations.

\begin{table}
\begin{center}
\begin{tabular*}{0.95\textwidth}{@{\extracolsep{\fill}}|llllll|}
\hline
$N_f$           & $10/g^2$ & $L^3\times T$ & $am_{\rm sea}$ & $am_{\rm val}$ & $N_{\rm cfg}$ \\
\hline
\hline
0         &   8.00 & $20^3\times 64$  & ---  & 0.020 & 408 \\
0         &   8.00 & $20^3\times 64$  & ---  & 0.050 & 408 $\longrightarrow$ 6154\\
\hline
2         &   7.20 & $20^3\times 64$  & 0.020 & 0.020 & 547 \\
\hline
2+1       &   6.76 & $20^3\times 64$  & 0.007, 0.05 & 0.007, 0.05 & 422\\
2+1       &   6.76 & $20^3\times 64$  & 0.010, 0.05 & 0.010, 0.05 & 644\\
2+1       &   6.85 & $20^3\times 64$  & 0.05, 0.05 & 0.05, 0.05 & 369\\
\hline
\end{tabular*}
\end{center}
\caption{Ensembles used for singlet calculations. 
\label{ensembles}}
\end{table}

We examined the $\gamma_5\otimes{\bf 1}$ taste-singlet state. This is generated
with a four-link operator $\Delta_{\gamma_5\otimes{\bf 1}}$
which displaces the quark and anti-quark sources
to opposite corners of the hypercube (and applies appropriate 
Kogut-Susskind phases). 
The parity partner of the 
$\gamma_5\otimes{\bf 1}$ state is exotic so correlators do not include 
contributions from an oscillating partner state. We use simple point 
sources to 
measure the connected correlators.

To measure the disconnected correlator we use stochastic Gaussian volume 
sources. Following Venkataraman and Kilcup \cite{Venkataraman:1997xi}, we 
use a noise-reduction trick that almost completely removes the component
of the variance that is due to the stochastic sources. The Venkataraman 
Kilcup variance reduction (VKVR) trick involves replacing the pseudoscalar
loop operator 
$\langle\eta\Delta_{\gamma_5\otimes{\bf 1}}M^{-1}\eta^\dagger\rangle_\eta$
with
$m\langle\eta\Delta_{\gamma_5\otimes{\bf 1}}M^{-1}M^{-1\dagger} \eta^\dagger\rangle_\eta$. 
The staggered $\Dslash$ property of connecting only odd to even lattice sites
ensures that the two have the same expectation value, but the latter has 
greatly decreased variance.

\section{Results}
\begin{figure}[tbh]
\resizebox{2.9in}{!}{\includegraphics{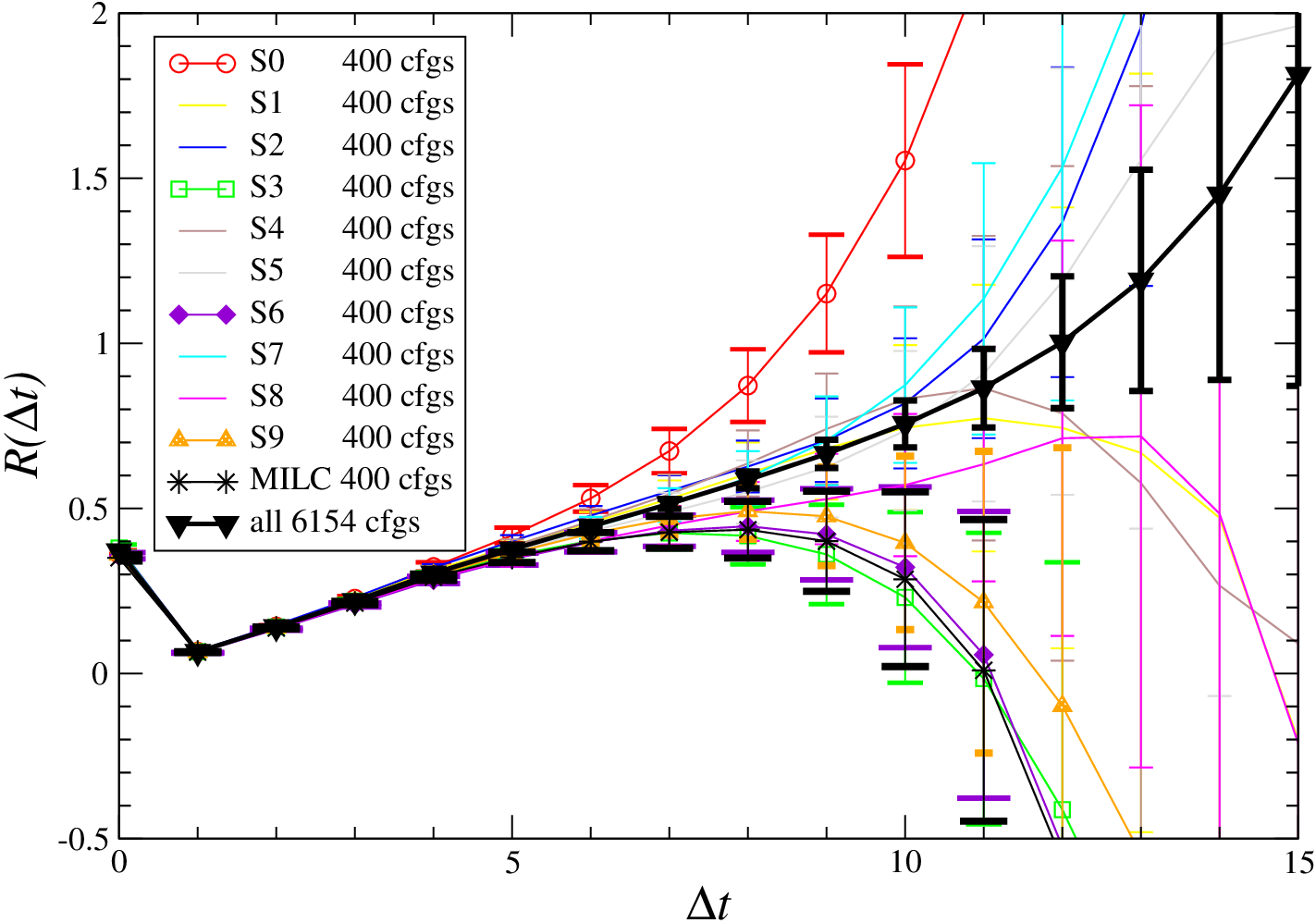}}
\resizebox{2.9in}{!}{\includegraphics{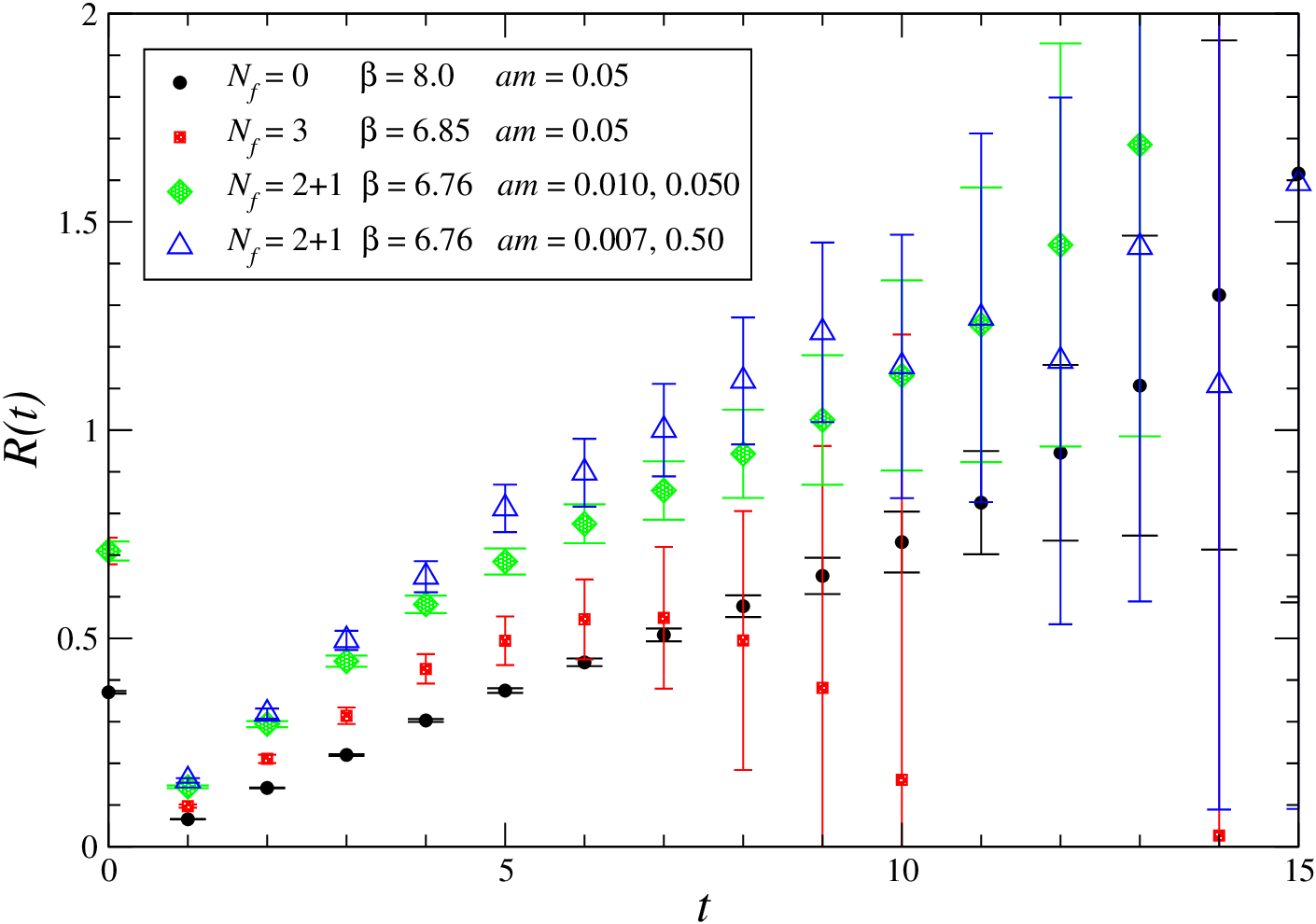}}
\caption{D/C ratio for 6154 $\beta=8.00$ quenched lattices with valence quark 
mass $am=0.05$, and for 11 subsets of 400 configurations. We highlight five of 
these subsets, including the original MILC configurations, as disagreeing with 
the mean by more than one $\sigma$.  \label{binned_dc_rat_b800}}

\caption{$D/C$ ratio for coarse MILC ensembles. For the $N_f=3$ and $N_f=2+1$
ensembles $R_{SU3}$ is displayed. For the quenched configurations the
single-flavour $R=D/C$ is plotted.}
\end{figure}

An initial calculation of the $D/C$ ratio (\ref{quenched_R}) on the 
408 quenched $\beta=8.00$ lattices with quark mass $am=0.05$
was not convincingly linear (bursts in 
Fig. \ref{binned_dc_rat_b800}). To explore why 
this was the case we extended the ensemble to 6154 configurations by 
generating more lattices in ten Monte Carlo streams and calculated the $D/C$ ratio on the full ensemble (bold inverted triangles in 
Fig. \ref{binned_dc_rat_b800}). With the increased 
statistics the $D/C$ ratio is now consistent with the form on 
(\ref{quenched_R}).

We separately extracted the non-singlet $\gamma_5\otimes{\bf 1}$
ground state from the connected correlator and found $M_{NP}=0.5180(3)$.
In doing a linear fit to the quenched $D/C$ ratio, and rescaling by 
$\sqrt{N_f}=\sqrt{3}$, we find $m_0= 0.76$GeV.

A similar calculation using the $R(t)_{SU3}$ ratio form in (\ref{fullQCD_R})
with the 644 $\beta=6.76$ configurations gives $a(M_{SP}-M_{NP})=0.24(9)$. 
From the connected correlator we get $aM_{NP}=0.571(2)$. Using $a=0.125$fm 
we extract $M_{NP}=1280(142)$MeV. 

\begin{figure}[tbh]
\resizebox{2.9in}{!}{\includegraphics{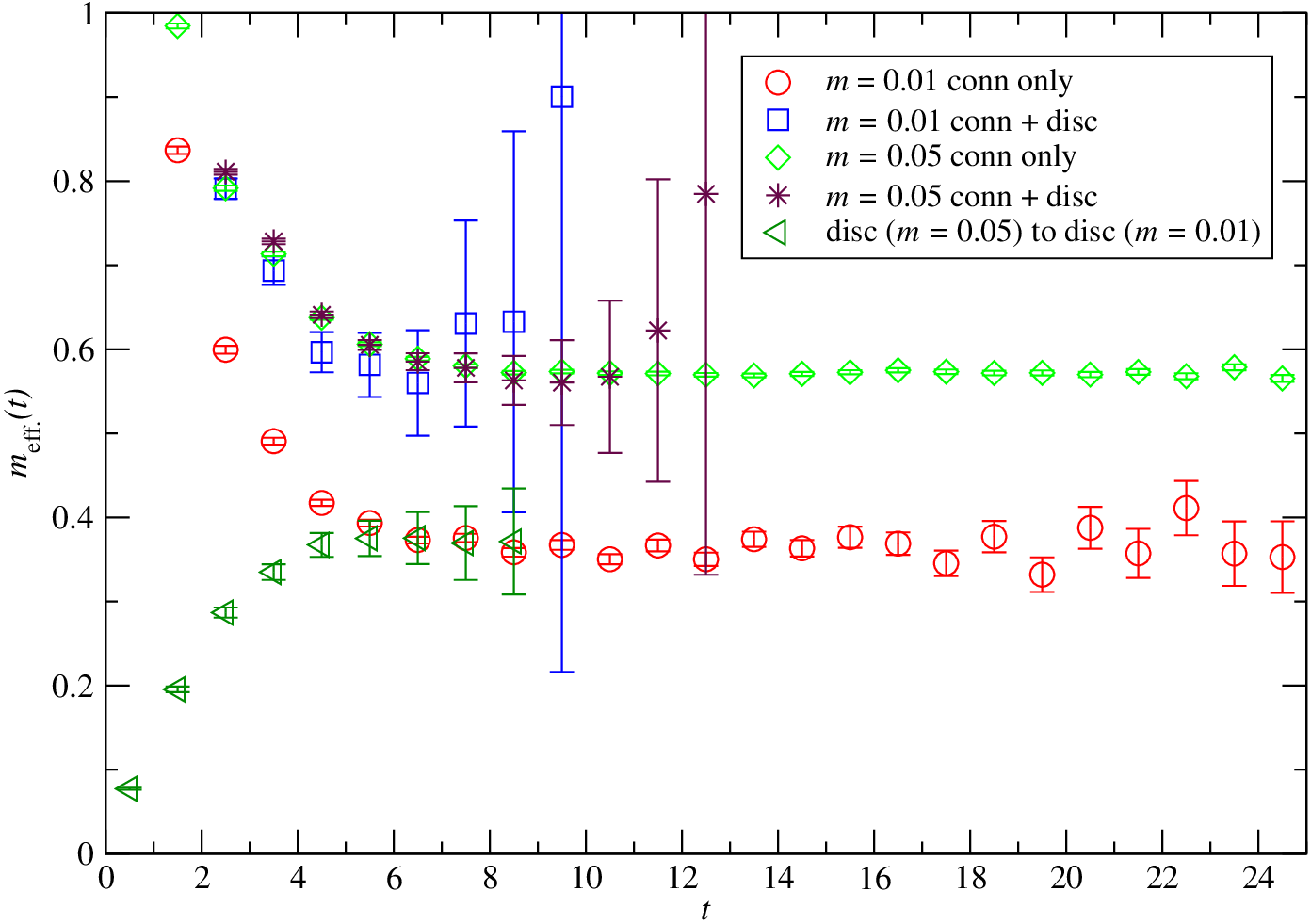}}
\resizebox{2.9in}{!}{\includegraphics{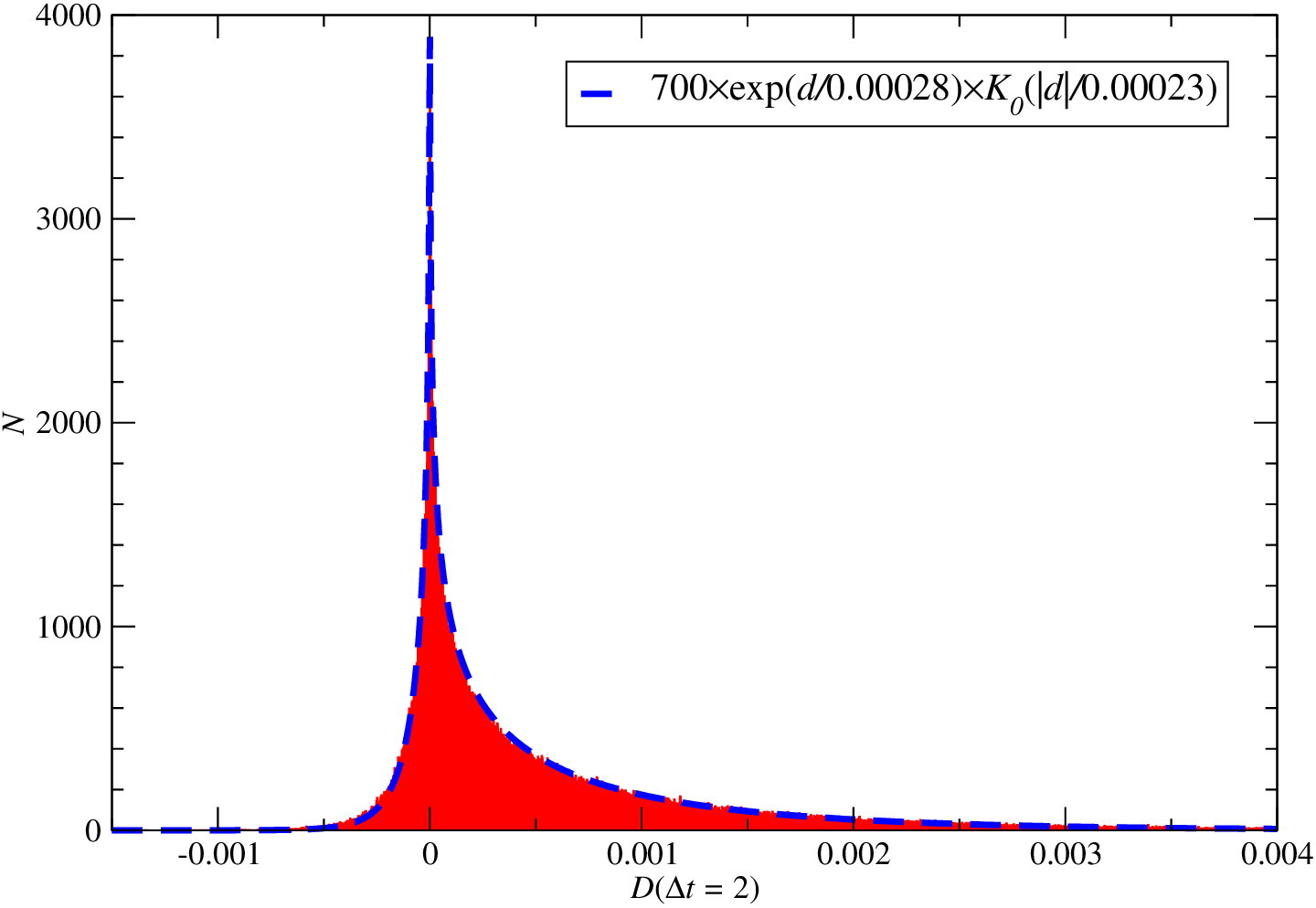}}
\caption{Effective masses for the $(\gamma_5 \otimes {\bf 1})$ channel
for the light and strange quarks 
for the $\beta$ = 6.76, m=0.01/0.05 data set.
\label{fig:meffB676}}
\caption{Histogram of 393856 $D(\Delta t=2)$ measurements on 6154 quenched 
lattices.
\label{g5_dcorr_histo_t2}}
\end{figure}

For the $\beta=6.76$, $am=0.01,0.05$ ensemble we also calculate the 
effective masses diagonal and off-diagonal elements of the variational 
propagator
\begin{equation}
{\bf G}(\Delta T)=
\left[ 
\begin{array}{cc}
{\bf C}_{qq}(\Delta t)-2{\bf D}_{qq}(\Delta t) &  -\sqrt{2}{\bf D}_{qs}(\Delta t) \\
-\sqrt{2}{\bf D}_{sq}(\Delta t) & {\bf C}_{ss}(\Delta t)-{\bf D}_{ss}(\Delta t)
\end{array} \right].
\end{equation}
These are shown in Fig. \ref{fig:meffB676}, along with effective masses of the 
light and strange connected correlators. The variational method relies on 
all elements of ${\bf G}(\Delta T)$ having the same ground state. We see in 
Fig. \ref{fig:meffB676}. that the diagonal elements do share a common ground 
state, corresponding to about $990$MeV. The off-diagonal elements of
${\bf G}(\Delta T)$ are mixed disconnected correlators, and the effective mass
corresponds to about $600$MeV. We would expect the lowest stable state in 
this channel to be the $\eta$ at $548$MeV. 

We suspect that the explanation for this discrepancy lies in the limited 
statistics available in these ensembles. 
The great difficulty in measuring $R(t)$ and the singlet propagators is in the 
fluctuations in the disconnected correlator. Unlike many observables in 
lattice simulations, disconnected correlator, as a product of two roughly
Gaussian-distributed quantities (the loop operators), has a distinctly 
non-Gaussian distribution, whose peak remains at zero. 
The mean value is determined by the asymmetry which, when the correlation is 
small, is almost entirely in the tails, many standard deviations away from the 
mean. An example is shown in Fig. \ref{g5_dcorr_histo_t2}. Our experience 
with the quenched ensembles shows that several hundred configurations is 
generally not sufficient to resolve the disconnected correlators to do 
meaningful spectroscopy in configurations of this size. We did not find
significant autocorrelations in any ensemble.

\section{Long dynamical ensemble}\label{long}
To attack the disconnected correlators forcefully, we have generated a 
$N_f=2+1$ flavor ensemble of 5081 configurations at intervals of six RHMC
trajectories. These $24^3\times 64$ configurations with $\beta=6.75$ and
$am=0.006,0.03$ have been generated on the UKQCD collaboration's QCDOC 
machine. We chose the strange quark simulation mass of $am_s=0.03$ to 
coincide with values used by MILC in newer coarse 
ensembles~\cite{Bernard:2005ei}.

\begin{table}
\begin{center}
\begin{tabular*}{0.95\textwidth}{@{\extracolsep{\fill}}|llllll|}
\hline
$N_f$           & $10/g^2$ & $L^3\times T$ & $am_{\rm sea}$ & $am_{\rm val}$ & $N_{\rm cfg}$ \\
\hline
\hline
2+1        &   6.75   & $24^3\times 64$  & 0.006, 0.030  & 00.006, 0.030 & 6000 \\
\hline
\end{tabular*}
\end{center}
\caption{Long run dynamical ensembles for singlet calculations. 
\label{longensembles}}
\end{table}

We have completed preliminary spectroscopy on a portion of this ensemble and 
find $aM_{\pi}=0.17478(35)$ and 
$aM_{\rho}=0.5298(17)$. A preliminary estimate of the static quark potential
gives $r_0/a=3.810(51)$.

\begin{figure}[tbh]
\resizebox{5.8in}{2in}{\includegraphics{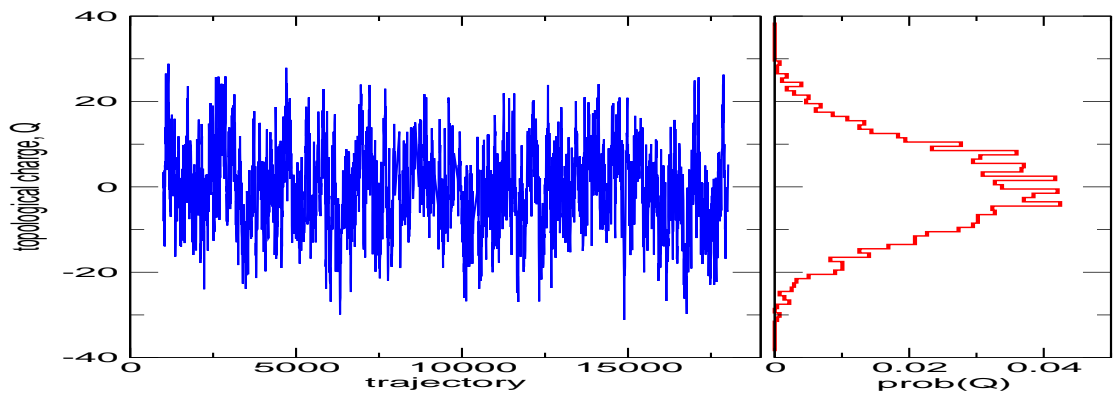}}
\caption{Topological charge time-series and histogram for about 17000 
trajectories of the long dynamical run.\label{longrunTOPO}}
\end{figure}

We have measured the topological charge on the ensemble 
and the time-history shows good tunnelling with an autocorrelation time
in the range of 30 - 50 trajectories (Fig.\ref{longrunTOPO}).

\section{Conclusions and Outlook}
We have completed the first phase of a project to explore singlet physics with 
staggered fermions. We have tested several methods of extracting singlet
masses from connected and disconnected correlator combinations.
We again refer the interested reader to the extended analysis in
\cite{Gregory:2007ev}, including
analysis of the other ensembles in Table \ref{ensembles}.

Comparing the original and the extended quenched ensemble highlights the 
difficulties of working with ``small'' ensembles, and the need for long 
dynamical-fermion ensembles to calculate disconnected 
diagrams for singlet physics.  We hope to be able to resolve the 
$\eta/\eta^\prime$ spectrum in the ensemble described in Section \ref{long}.

\section{Acknowledgements}        
A portion of this analysis was performed on ScotGrid.


\begin{thebibliography}{99}
\bibitem{Witten:1979vv}
E.~Witten,
\newblock Nucl. Phys. {\bf B156}, 269 (1979),
\newblock 

\bibitem{Veneziano:1979ec}
G.~Veneziano,
\newblock Nucl. Phys. {\bf B159}, 213 (1979),
\newblock 

\bibitem{Itoh:1987iy}
S.~Itoh, Y.~Iwasaki, and T.~Yoshie,
\newblock Phys. Rev. {\bf D36}, 527 (1987),
\newblock 

\bibitem{McNeile:2000hf}
UKQCD, C.~McNeile and C.~Michael,
\newblock Phys. Lett. {\bf B491}, 123 (2000), hep-lat/0006020,
\newblock 

\bibitem{Struckmann:2000bt}
TXL, T.~Struckmann {\em et~al.},
\newblock Phys. Rev. {\bf D63}, 074503 (2001), hep-lat/0010005,
\newblock 

\bibitem{Lesk:2002gd}
CP-PACS, V.~I. Lesk {\em et~al.},
\newblock Phys. Rev. {\bf D67}, 074503 (2003), hep-lat/0211040,
\newblock 

\bibitem{Schilling:2004kg}
K.~Schilling, H.~Neff, and T.~Lippert,
\newblock Lect. Notes Phys. {\bf 663}, 147 (2005), hep-lat/0401005,
\newblock 

\bibitem{DeGrand:2002gm}
MILC, T.~A. DeGrand and U.~M. Heller,
\newblock Phys. Rev. {\bf D65}, 114501 (2002), hep-lat/0202001,
\newblock 

\bibitem{Venkataraman:1997xi}
L.~Venkataraman and G.~Kilcup,
\newblock (1997), hep-lat/9711006,
\newblock 

\bibitem{Kogut:1998rh}
J.~B. Kogut, J.~F. Lagae, and D.~K. Sinclair,
\newblock Phys. Rev. {\bf D58}, 054504 (1998), hep-lat/9801020,
\newblock 

\bibitem{Fukaya:2004kp}
H.~Fukaya and T.~Onogi,
\newblock Phys. Rev. {\bf D70}, 054508 (2004), hep-lat/0403024,
\newblock 

\bibitem{Aoki:2006xk}
JLQCD, S.~Aoki {\em et~al.},
\newblock (2006), hep-lat/0610021,
\newblock 

\bibitem{Gregory:2007ev}
  E.~B.~Gregory, A.~C.~Irving, C.~M.~Richards and C.~McNeile, (2007)
  arXiv:0709.4224 [hep-lat].


\bibitem{Davies:2003ik}
HPQCD, C.~T.~H. Davies {\em et~al.},
\newblock Phys. Rev. Lett. {\bf 92}, 022001 (2004), hep-lat/0304004,
\newblock 

\bibitem{Aubin:2004wf}
C.~Aubin {\em et~al.},
\newblock Phys. Rev. {\bf D70}, 094505 (2004), hep-lat/0402030,
\newblock 

\bibitem{Durr:2005ax}
S.~Durr,
\newblock PoS {\bf LAT2005}, 021 (2006), hep-lat/0509026,
\newblock 


\bibitem{Sharpe:2006re}
S.~R. Sharpe,
\newblock PoS {\bf LAT2006}, 022 (2006), hep-lat/0610094,
\newblock 


\bibitem{Creutz:2007yg}
M.~Creutz,
\newblock Phys. Lett. {\bf B649}, 230 (2007), hep-lat/0701018,
\newblock 

\bibitem{Creutz:2007rk}
M.~Creutz,
\newblock (2007), arXiv:0708.1295 [hep-lat],
\newblock 

\bibitem{Bernard:1992mk}
C.~W. Bernard and M.~F.~L. Golterman,
\newblock Phys. Rev. {\bf D46}, 853 (1992), hep-lat/9204007,
\newblock 

\bibitem{Bernard:2005ei}
  C.~Bernard {\it et al.}  [MILC Collaboration],
  PoS {\bf LAT2005}, 025 (2006)
  [arXiv:hep-lat/0509137].

\end{thebibliography}
\end{document}